\documentclass{sig-alternate-05-2015}
\usepackage{graphicx}
\usepackage{enumerate}

\begin{document}

\setcopyright{acmcopyright}

\title{Development of a Greek language teaching platform for primary school pupils}

\numberofauthors{1} 
%
\author{
%
%
\alignauthor
Eleni Michailidi, Ioannis Skordas, Maria Papatsimouli, Lazaros Lazaridis, \\Heracles Michailidis, Stavroula Tavoultzidou and George F. Fragulis \\
       \affaddr{Lab. of Web Technologies \\ 
\& Applied Control Systems}\\
       \affaddr{Dept. of Electrical and Computer Engineering}\\
       \affaddr{University of Western Macedonia}\\
       \affaddr{Koila, Kozani}\\
       \affaddr{Hellas}\\
       \email{gfragulis@uowm.gr}}


\maketitle
\begin{abstract}
 This study aims to develop an online Greek language learning platform via a learning management system based on the WordPress platform for primary school pupils with Russian as a mother tongue. The subject was chosen with a view to aid both the communication needs of Russian native speakers in Greece, as well as their smooth integration into society. For the purpose of the current study, we investigated how Greek language is learnt in Russia, how widespread and popular it is, whether it is supported by the Russian educational system, as well as whether Greek language distance learning could be achieved. Moreover, LMS(Learning management system) environments were investigated as well, aiming at selecting the most student/teacher-friendly one. Taking into account that the platform is addressed to children, a presentable and properly developed material was designed, thus making the learning of each module more interesting and comprehensible throughout its short duration. With "GrfKids" platform speakers of other languages have the opportunity to learn Greek in an interactive asynchronous teaching environment with an automatic assessment. Moreover, "GrfKids" can be easily and immediately expanded both to pupils` linguistic, as well as age range. "GrfKids" provides Greek language courses to children with Russian as a mother tongue with interactive modules in which the learners can adapt the learning time by themselves. It is easy to use, functional and easily adapted to the ages it is addressed to. It was developed to create a teaching system by applying different plugins to the selected CMS(Content Management system), aiming to achieve a curriculum in which the pupil is in direct interaction with the educational material. It can be used individually, or as a support tool in a real class context. The curriculum has been designed in an evolutionary format and there is an introduction in each module explaining the part that follows.

\end{abstract}

%
%


%
%

%
%



 \section{Introduction}
 
 It is generally accepted that within a globalized society in which the use of foreign languages has increased, foreign languages are a prerequisite for communication \cite{Mehisto2012}. Moreover, proficiency in more than one language can enrich academic and life experiences, while it can also equip with the necessary qualifications for personal and career advancement. Teaching a second /foreign language, as well second language acquisition have always been characterized by searching for new effective ways of teaching/learning a new language on the one hand and the adaption of new methods and approaches on the other. Thus, foreign language education entails a variety of methods and approaches aiming at fulfilling the various learners needs, as well as making learning process more permanent and effective \cite{Simsek2019}, \cite{Wang2017}.  Each approach or method can be distinguished by its theory and sets of principles as to how language is best taught and learned \cite{Czasny2019}.

 The use of computers for foreign language learning is a process with numerous benefits, i.e., increasing students attention to the subject area, creating incentives to improve writing etc \cite{Blake2013}. In a society of knowledge in which information is transmitted over the Internet at a rapid speed, the school has to prepare pupils as the citizens of tomorrow to join this new social model \cite{Ackermann2014}, \cite{Caladine2008}, \cite{Welsh2003}.  In the new generation of the Web (Web 2.0), users interact, collaborate and shape the content of web pages altogether. Nowadays, a user can learn everything on the internet, i.e. from solving a query to adding qualifications to a CV. The main advantage of this constant flow of information and knowledge is that every user can choose the way he/she wishes to learn \cite{Alexander2006}.  

Within this framework, there are several online platforms for learning Greek and foreign languages that are, either free or pay ones. However, most of them are, either static pages or with a question/answer format lacking a full description of every module. Furthermore, while Greek language teaching is widespread, there is no specific educational content management system for teaching the Greek language to children with Russian as a mother tongue. Some of the best-known learning platforms for Greek language learning are:

\begin{itemize}
 \item Duolingo, a language learning platform allowing the user to choose the language he/she wants to learn at his/her own pace, while he/she can also contribute to the expansion of the curriculum \cite{Duolingosupport2019}.
 \item Loecsen, a language learning system involving audiovisual educational material, which, however, is not interactive \cite{Loecsen2019}.  
 \item Ilearngreek.com, a Greek language learning website involving text reading and audio material \cite{Ilearngreek2019}.
\end{itemize}

Taking into account all the above mentioned, the present study presents the "GrfKids" application, accessible online (http://www.greekforkids.org), which aims at fulfilling the Russian tourists visiting Greece communication needs, as well as teaching Greek language to economic migrants children in order to be integrated into the school environment as soon as possible. It is worth pointing out that "GrfKids" platform can be used entirely by children, without the support of an adult/ Greek language speaker. Finally, it can be used in a real class context, as a support material for the teacher himself/herself.The platform was developed in WordPress, an open-source content management system based on PHP and MySQL, allowing to upload and manage web content on the web \cite{Avgeriou2003}, \cite{Brazell2011}, \cite{Powers2019}, \cite{Schwartz2012}. It has many features including a plugin architecture and a template system. Wordpress is among the most popular systems in use on the Internet. Wordpress is among the most popular systems in use on the Internet, since it can ensure security and easy-to-use programming software both for developers and website administrators  \cite{Koskinen2012}, \cite{Krol2019}, \cite{Williams2011}. WordPress plugins provide additional functions \cite{Hills2016}, \cite{Hrastinski2008},  and are used in Learning Management Systems \cite{Lonn2009} that supply organized training material in a structured format and formed courses in an evolutionary format, usually created with web tools.

A number of application results using other open source programming
languages such as PHP and MySQL can be found in \cite{Fragulis2018}, \cite{Lazaridis2019},\cite{Skordas2014}, \cite{Skordas2017} and the references herein. For image, animation and video editing there were used: a) Adobe Photoshop, an image editing program; b) Adobe Flash, now called Animate for animation, multimedia, applications, mobile games, and c) Adobe Premiere for video editing .

 \section{Characteristics of the proposed Web Platform}
 
 \subsection{Advantages of using Web Technologies in Education and Distance Learning}
 The proposed web-based environmental application is accessible, via any personal computer or a smartphone/tablet with an Internet connection, everywhere at anytime. Hence, the user can retrieve the information he/she needs from the online platform easily and quickly \cite{Shaffi2013}. In addition, the user interface of web-based applications is easier to customize than in other types of applications. Therefore, the administrator can update the GUI of the application or can customize the presentation of information to different user groups.

 \subsection{Analysis of various issues during development phase}
 
In order to evaluate the usability, functionality, and effectiveness of the "GrfKids" Web application, a number of steps took place following the process described below:

\begin{itemize}
 \item The application was uploaded in a web server and the lessons were made available to the test-users. 
 
 \item The speed/response of the application was tested. The results were very satisfactory, as the application worked rather fast.

 \item An account for every user type (Pupils /Teachers /Admins) was created. Users coming from different backgrounds (system developers, computer analysts, school teachers, Department of Primary Education researchers, Primary Education Universities students, and finally a group of newspaper/web-journalists) were tested, in an attempt to involve various categories of users with differences in age, level of education and occupation. 

 \item The test-users feedback was taken into account and helped us improve several parts of the application.  Specifically, during the alpha and beta test, the users pointed out that the functionality of the site, i.e. lessons, videos, and animation images, should be improved.

 \item All in all, the final version of the web system satisfied the majority of the test-users, more specifically, the changes of the front-end of the application, as well as the variety of information provided. Deliberately little or no information was provided to the test-users about the application of the web platform in order to test whether the platform is functional and easy to use, i.e., a number of pupils at home/ or in a classroom , or a group of tourists visiting Greece. However, the majority of the comments were positive and encouraging and the test-users found the application easy to handle and use.
\end{itemize}

 \subsection{Functional \& non-Functional requirements}
 
The functional and non-functional requirements of the platform are:\\
\textbf{Functional}:
\begin{itemize}
 \item Easy User registration.
 \item The platform supports three types of users:
  \begin{itemize}
    \item The teacher who has the ability to define the educational material and check the pupils progress.
    \item The pupils who have access to the teaching material.
    \item The administrator, who has full access to every part of the platform.
    \end{itemize}
 \item The platform provides tools to create and configure digital interactive educational content
\end{itemize}

\textbf{Non-functional}:
\begin{itemize}
 \item It is available online and it is supported by every operating system.
 \item There is no age barrier. It can be used by very young pupils.
 \item It requires no installation.
 \item It is supported by mobile devices.
 \item It is presentable.
\end{itemize}

 \subsection{User groups}
  
\begin{figure}[h!]
\centering
\includegraphics[scale=0.42]{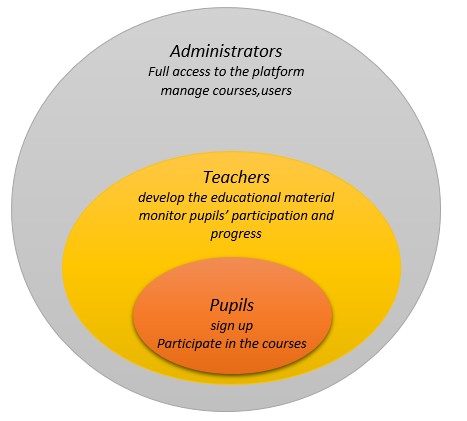}
\caption{Flow diagram of the "GrfKids" system}
\label{fig:f1}
\end{figure}

 The system diagram presented in Figure \ref{fig:f1} describes the user groups of the platform. The first group is the "pupil", who can sign up, as well as participate in the courses in an interactive way at his/her own pace and get an automatic assessment. The second group is the "teacher", who can develop the educational material and monitor pupils participation and progress. The third group is the "administrator" who has full access to every part of the system and thus the ability to manage courses and users of the two groups above mentioned.

\subsection{Description of the system}

For the deployment of the application, the Raspberry Pi (RPi) system was selected, a low-cost single-board computer. It was configured as a network server, in order to examine the effectiveness of the application in providing teaching via wireless local area network at areas without internet access and without the use of mobile networks, with only pre-requisite for the user a wireless network connection capable device with a browser program.  Such environments may be the environment of a class or a tourist information kiosk.

To create the content of the lessons, images were used for a better configuration of the learning material of the platform as a whole. These images were presented in a single format or per video group. For better results, it was necessary to use editing tools to format the graphic content including creation and/or editing of images, videos, and effects. Thus, Adobe Photoshop, Adobe Flash and Adobe Premiere were used for this purpose. More specifically, Photoshop for editing and customizing graphics in the teaching material of the platform. Graphics were adjusted for size, color, and background. Also, there was a combination of graphics to create a single image. Flash made the animation used in the videos. In addition, styles created writing effects and image motion on a fixed background. Finally, Premiere processed the videos that were used thus creating more dynamic content. The material resulting from image, animation, and video edits was used throughout the platform.

For a better appearance and functionality of the page, the web platform used Smart Slider3, Page Builder, Shop Page WP, and Wordpress Ultimate Member plugins \cite{plugins2019}. 

The web application also used H5P, a free and open-source JavaScript-based framework that helps users create, share and reuse interactive HTML5 content \cite{H5P2019}. Based on this add-on, all platform modules can be formatted with multiple choice questions, right image selection, cards, memory games, and spelling. H5P is used in conjunction with xAPI to store data, as well as content transfer to another management system .

Krashen student-centered learning theory constituted the basis for content modifications without significant variations. According to Krashen, lessons are designed to guide and assist with the smooth outcome of all content and not with the absolute and strict compliance with it \cite{Krashen1981}, \cite{Krashen1982}.

\section{User Interface and Functionality}

In this part, there is a detailed description of the "GrfKids" application interface and features available to the user.

\begin{figure}[t]
\centering
\includegraphics[scale=0.38]{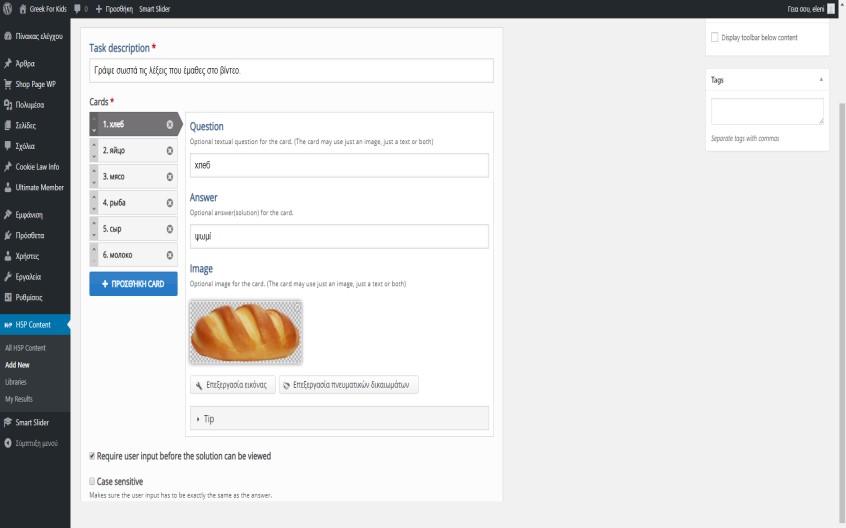}
\caption{H5P add-on}
\label{fig:f2}
\end{figure}

According to Figure \ref{fig:f2}, the user with administrator/ teacher rights has the ability to customize the H5P plugin for modules configuration in order to format the way in which they are presented, such as multiple choice answers, selecting the correct image, memory and spelling games.

\begin{figure}[t]
\centering
\includegraphics[scale=0.35]{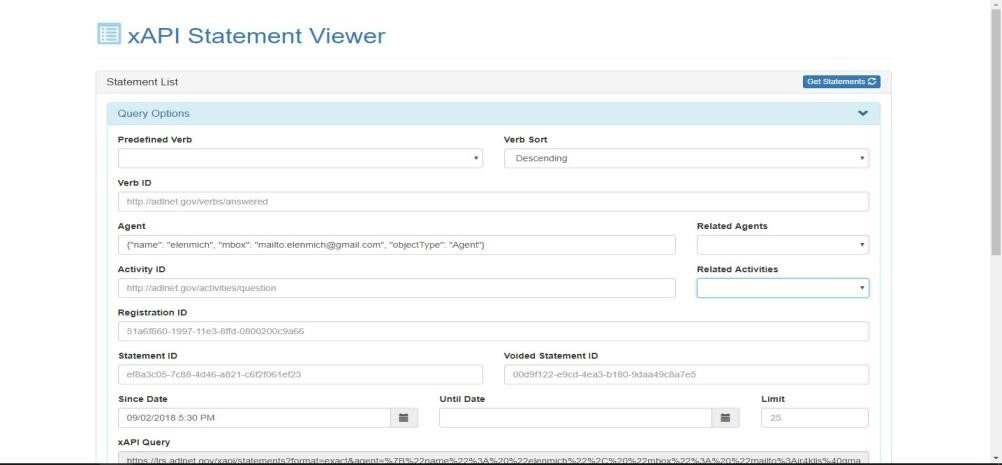}
\caption{Xapi plugin}
\label{fig:f3}
\end{figure}

As shown in Figure \ref{fig:f3}, the user can use the xAPI plugin together with H5P for data storage. This combination transfers and modifies files that are secured because it works not only through the platform itself but also within the H5P itself, where the files are kept and can be transferred if the content is transferred to other management systems.

\begin{figure}
\centering
\includegraphics[scale=0.35]{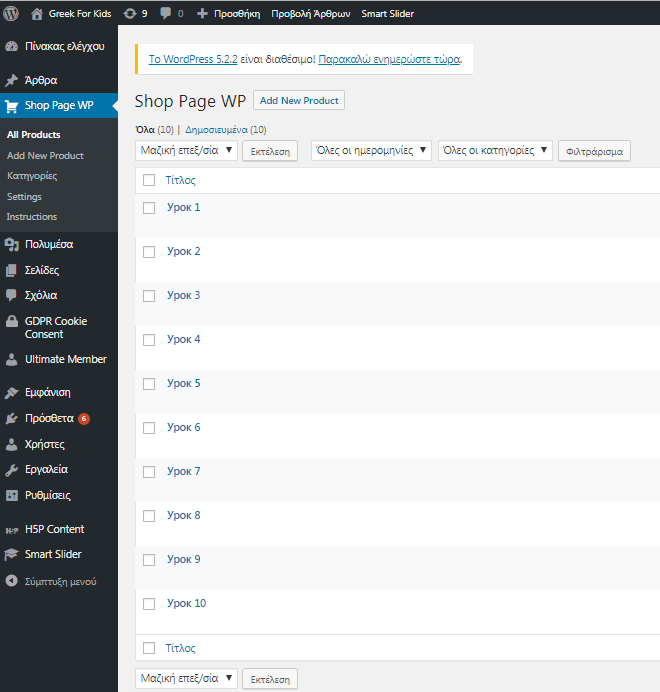}
\caption{Creating, editing and deleting a course page (Shop Page WP)}
\label{fig:f4}
\end{figure}

Shop Page WP is a plugin used to create, edit, and delete the course page (Figure \ref{fig:f4}).

\begin{figure}
\centering
\includegraphics[scale=0.20]{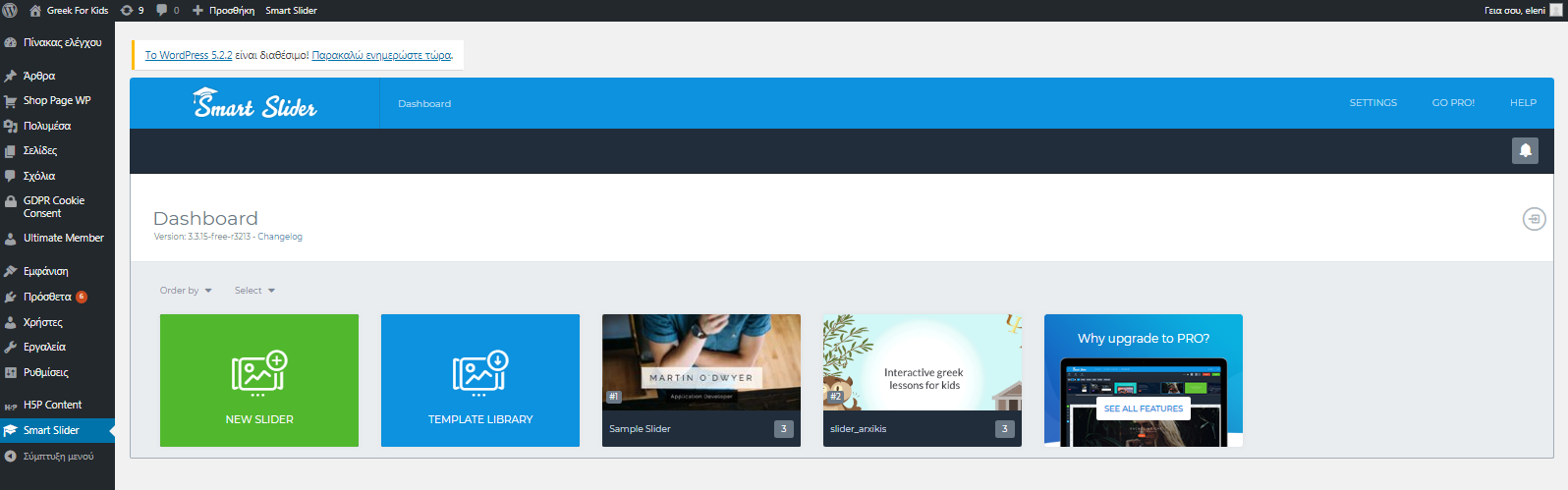}
\caption{Using Smart Slider 3, for "GrfKids" home screen}
\label{fig:f5}
\end{figure}

As shown in Figure \ref{fig:f5}, Smart Slider 3 is the most powerful and intuitive WordPress plugin for settings that could not be done earlier. It is very easy to edit, compatible with SEO optimizations, as well as with most WordPress themes. In this case, it was used on the home screen and integrated with PHP.

\begin{figure}[h!]
\centering
\includegraphics[scale=0.35]{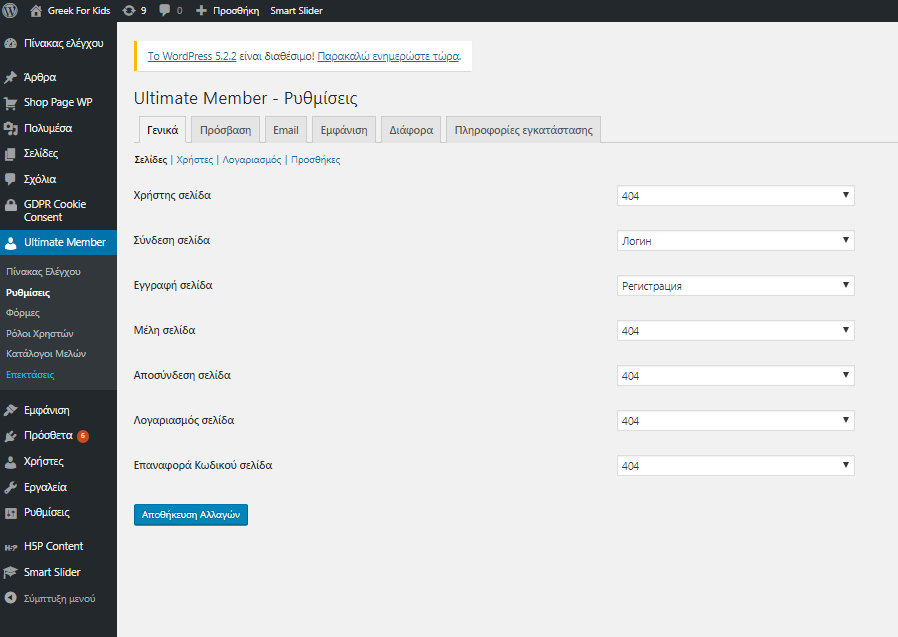}
\caption{Using Ultimate Member for members sign up}
\label{fig:f6}
\end{figure}

Ultimate Member plugin is used for creating the users profile and the addition of a new member to WordPress. This plugin is a handy choice for users to sign up and become a member of a website, while it also allows the addition of visible user profiles to the site. In addition, it is ideal for creating advanced online communities and access sites. It is functionally lightweight and very flexible and it also enables the creation of almost any kind of site, where users can sign up with great ease (Figure \ref{fig:f6}).

\begin{figure}[h!]
\centering
\includegraphics[scale=0.21]{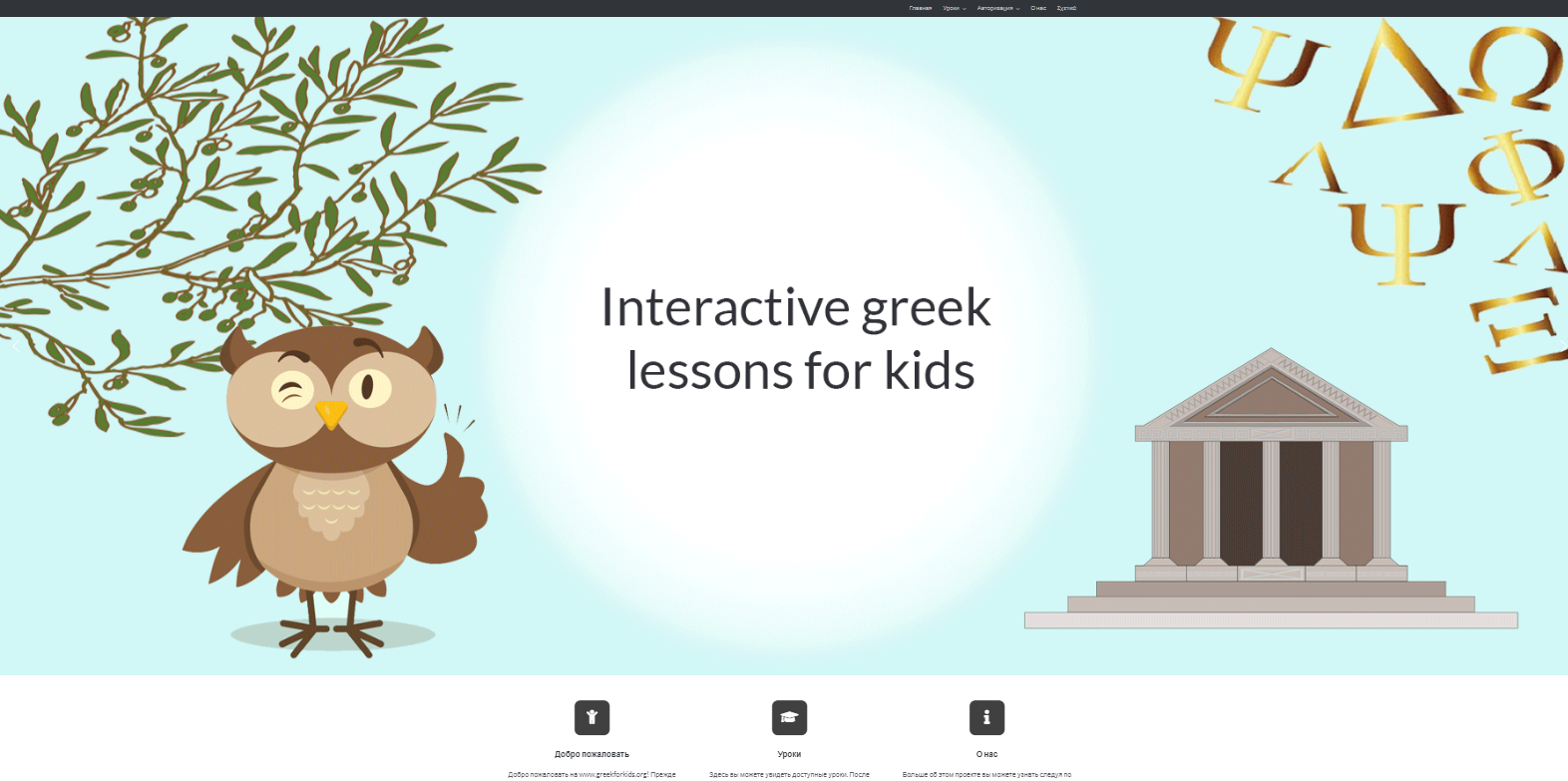}
\caption{"GrfKids" homepage}
\label{fig:f7}
\end{figure}

All the above tools are used for the LMS "Greek for kids" page development. Only the key points and not every step and detail of the development process have been presented here.
What follows are descriptions concerning the users visit to our page. All the functions, as it has been mentioned before, are in Russian, due to the purpose the platform was built for.
Figure \ref{fig:f7} shows the home page of the web application with the corresponding menus, i.e. "lessons", "sign up", "about us" while the option for cookies acceptance is also displayed at the bottom of the screen.

\begin{figure}
\centering
\includegraphics[scale=0.35]{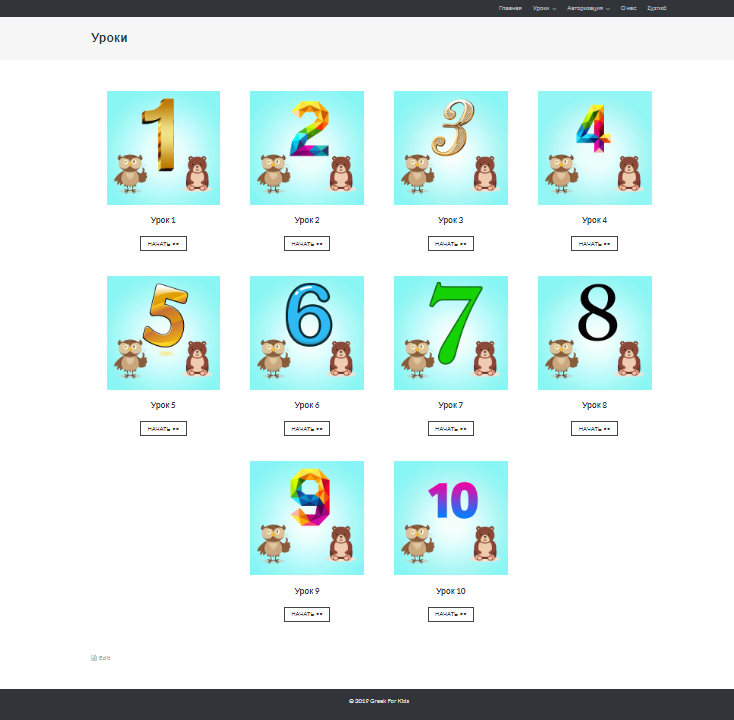}
\caption{ Web platform course options}
\label{fig:f8}
\end{figure}

By selecting the lessons menu, icons representing each lesson are displayed. By selecting the button located at the bottom of each icon, the user is redirected to the corresponding lesson (Figure \ref{fig:f8}).

\begin{figure}[b]
\centering
\includegraphics[scale=0.35]{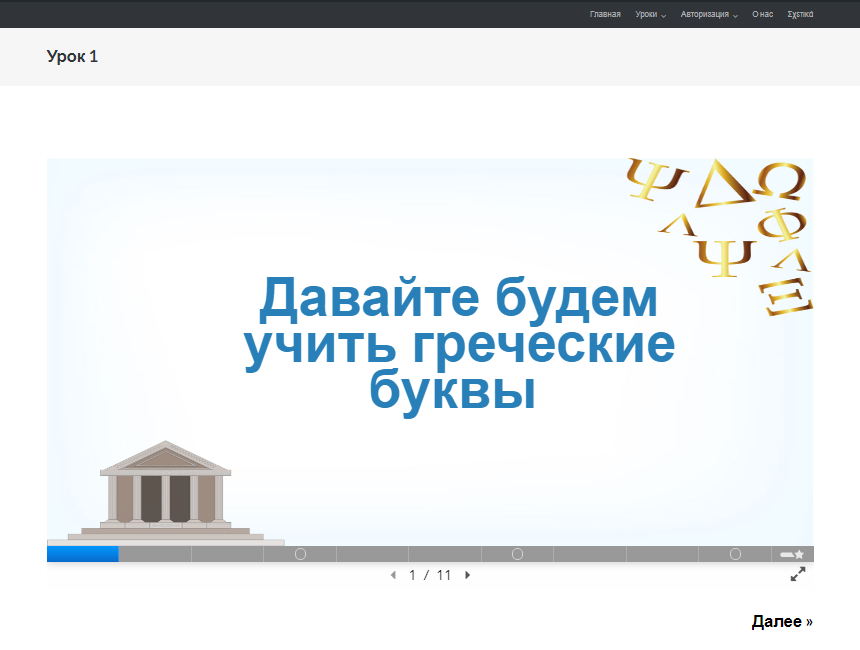}
\caption{Display of the first Lesson}
\label{fig:f9}
\end{figure}

In Figure \ref{fig:f9} the first lesson is presented. There are images, videos, games, all embedded in a single application. At the bottom there is the option next, referring to the second lesson.

\begin{figure}[t]
\centering
\includegraphics[scale=0.35]{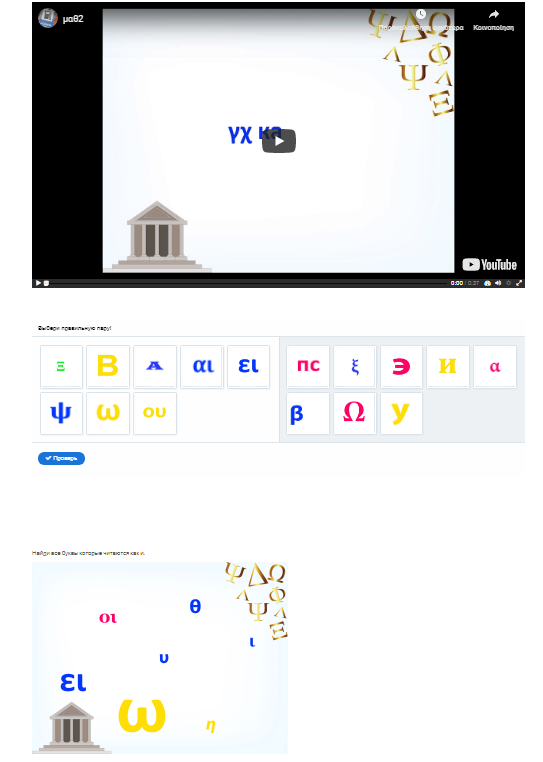}
\caption{Lesson with an introductory video and interactive games}
\label{fig:f10}
\end{figure}

In the second lesson, there is an explanatory video which constituted the basis for the creation of two interactive games, i.e., matching similar objects and matching the correct answer in a context with active points. (Figure \ref{fig:f10}).

\begin{figure}[b]
\centering
\includegraphics[scale=0.35]{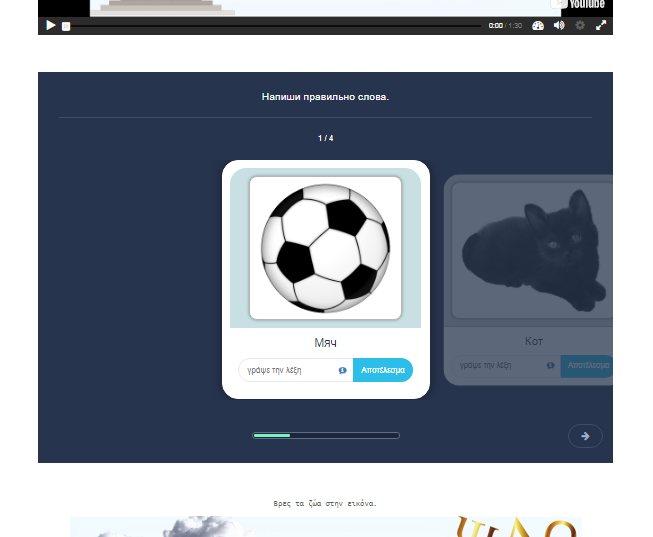}
\caption{A lesson with a spelling game}
\label{fig:f11}
\end{figure}

Lesson 3 is structured in a similar way as the previous one, the only difference being a spelling game of the words taught in the introductory video (Figure \ref{fig:f11}).

\begin{figure}[t]
\centering
\includegraphics[scale=0.35]{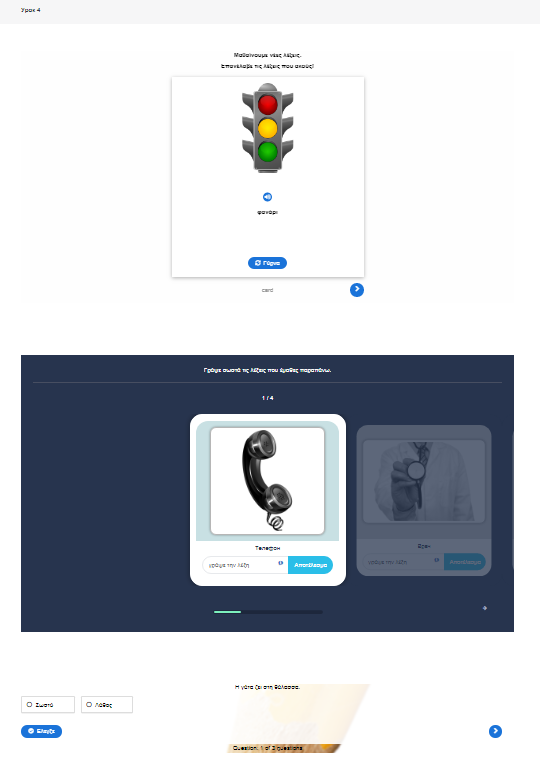}
\caption{Lesson using audio cards  }
\label{fig:f12}
\end{figure}

In lesson 4, new words are taught through audio cards translated into Russian for a better understanding of the meaning. The exercises involved are spelling, true/false questions and multiple choice answers (Figure \ref{fig:f12}).

\begin{figure}[b]
\centering
\includegraphics[scale=0.25]{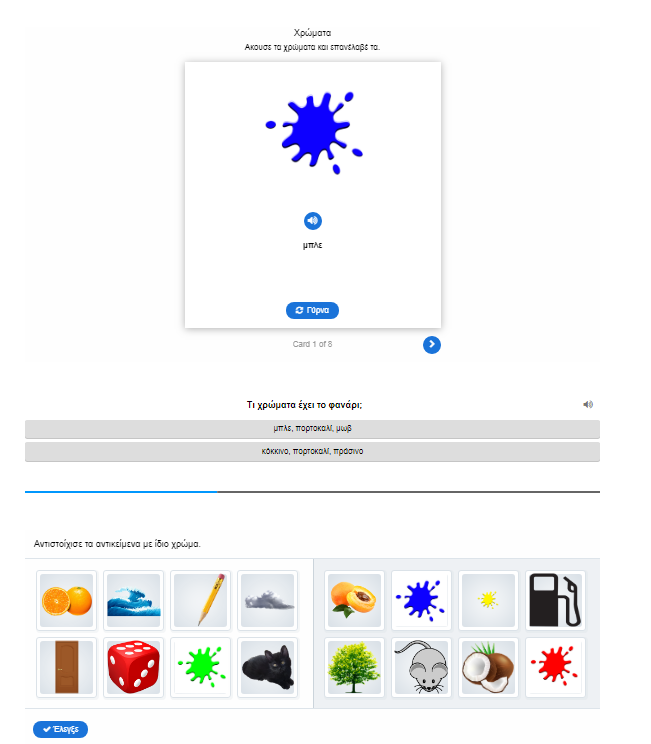}
\caption{Lesson using audio cards (cont. }
\label{fig:f13}
\end{figure}

In the fifth lesson, new vocabulary is taught in a similar way as before, while the exercises are also similar to the exercises of the previous lessons (Figure \ref{fig:f13}).

\begin{figure}[t]
\centering
\includegraphics[scale=0.25]{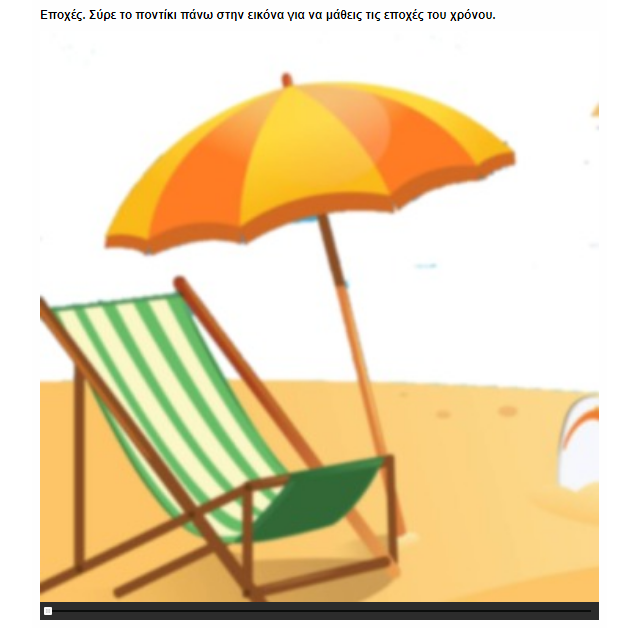}
\caption{A lesson about the seasons of the year}
\label{fig:f14}
\end{figure}

In lesson six, the seasons of the year are taught, as shown in the exercise above. By placing the cursor on the icon, there is also a change of the text (Figure \ref{fig:f14}).

\begin{figure}[b]
\centering
\includegraphics[scale=0.35]{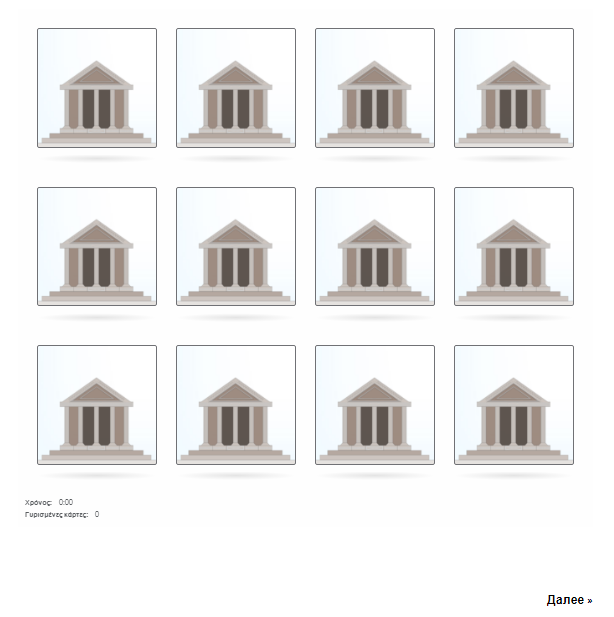}
\caption{Lesson using memory cards}
\label{fig:f15}
\end{figure}

In the seventh lesson, apart from similar exercises to previous lessons, there is an exercise with memory cards, i.e., by turning similar cards upside down, the name of the object is displayed (Figure \ref{fig:f15}).

Lessons, eight, nine, and ten are structured in the same way as the previous ones with only the content being different each time.

\section{Comparison of "GrfKids" with other Language Platforms-Discussion}

A number of similar platforms (Duolingo, Loecsen and Ilearngreek.com for teaching/learning a foreign language) to "GrfKids" platform have been presented in the Introduction. When comparing "GrfKids" platform with them, the following similarities and differences can be observed:

\textbf{Similarities}
\begin{itemize}
 \item It is free.
 \item It provides learning modules configured by the teacher-manager, in a similar way as Loecsen and ilearngreek.com.
 \item It has interactive courses with an automatic assessment like Duolingo.
 \item The learning modules are designed by the platform manager. 
\end{itemize}

\textbf{Differences}
\begin{itemize}
 \item It is built on an open source philosophy.
 \item It has interactive lessons with a preceding video explaining each lesson.
 \item It is addressed to a specific target audience.
\end{itemize}

\subsection{Benefits of the proposed system}
 As far as the benefits of "GrfKids" learning platform are concerned, while compared with the other Greek language learning sites above mentioned, it is worth pointing out that "GrfKids" can provide an easy-to-use educational program for children with the following characteristics:

 \begin{enumerate}[i]
 \item Immediate access to courses with signing up.
 \item A full explanation of each module.
 \item Interactive courses with an automatic assessment.
 \item Ability to adapt time according to each pupils needs.
 \item Courses designed on an evolutionary format.
 \item Possibility to expand and enrich the existing material or add new.
\end{enumerate}

\subsection{Limitations}
Concerning the constraints of the present platform, it is questionable whether consistency will be maintained by the student in an out of class environment, while the learning time cannot be defined accurately as well. However, a ten-day period may be set (for the existing material) provided that five minutes are given per day. Pupil data is recorded on the platform after signing up with the Ultimate member. Each lesson assessment results are recorded in the H5P add-on.

\section{Conclusions}

The study as a whole entails a combination of functionality and plugins, while research on the existing technologies, how they work and respond to each other, has been made as well. 

At first, the content of the syllabus was decided. Ritova book "Ta Ellinika"(Greek Language) \cite{Ritova1978}, \cite{Ritova2004} was chosen as a source for the ten lessons created on the platform. However, it should be noted, that the teaching material was developed more concisely than in conventional teaching of a second language since the focus of the present study was rather on developing a learning management system than on teaching. In order to teach the specific syllabus within the platform, several theoretical approaches were studied with Krashen approach \cite{Krashen1981}, \cite{Krashen1982} chosen as the most appropriate one. To create the curriculum modules in the WordPress management system, the  H5P plugin was used, approved by the Ministry of Education, Research and Lifelong Learning for the project "Digital School II Action 1" developed by the CTI  "DIOPHANTUS". The H5P plugin is an appropriate and easy-to-use option for syllabus design involving a variety of choices that respond to most theoretical approaches for material development. 
After completing the curriculum configuration, the WordPress system was modified to the LMS specifications. For this purpose, several available plugins were tested. However, since some of them could not function properly in conjunction with the H5P plugin, this process proved to be quite complex. Finally, the Ultimate Member plugin was implemented, which with some modifications, could work for pupil signing up and platform use.

The limits set in the research and the implementation of the study concerned the extension of the page and its application in real conditions. In particular, the translation of the material into other languages and the application in a real class context were confined. Moreover, a defined and limited number of lessons was set in order to implement the syllabus within the given definition. This is something that can be investigated in the future, as well as some other issues related to this study. In detail, the proposals for further research are the following:

\begin{enumerate}
 \item Expansion of the platform in terms of teaching Greek to pupils native speakers of other languages besides Russian i.e. Japanese, Chinese or Arabic.
  \item Applying the material to a real class learning context in the form it is presented, and recording statistics for learning outcomes in pupils progress during the e-learning process.
   \item Expansion of the platform teaching material.
    \item Application of the platform teaching material in a real class learning context using Raspberry Pi or any other smart devices.
\end{enumerate}

\section{Acknowledgments}
Dr. George F. Fragulis' work is supported by the no.80289 ELKE program "Educational Assesment System for Distance Learning" of the University of Western Macedonia, Kozani, Hellas.

%
\bibliographystyle{abbrv}

\end{document}